\documentclass[prb,amsfonts,amssymb,floats,twocolumn,superscriptaddress,aps]{revtex4-2}
\renewcommand{\paragraph}[1]{\textit{#1.}\textemdash}
\newcommand{\ui}{\mathrm{i}}

\makeatletter
\renewcommand{\fnum@figure}{\hspace{11pt}FIG. \thefigure}
\makeatother

\usepackage{comment} 
\usepackage{graphicx}
\usepackage{pythontex}
\usepackage{bm}
\usepackage{amsmath}
\usepackage[utf8]{inputenc}
\usepackage{amssymb}
\usepackage{dsfont}
\usepackage{mathrsfs}
\usepackage[english]{babel} 
\usepackage{enumitem}
\usepackage{braket} 
\usepackage{bbold}  
\usepackage{listings}
\usepackage{booktabs} 
\usepackage[colorlinks=true, breaklinks=true, linkcolor=blue, citecolor=blue, urlcolor=blue]{hyperref} 
\usepackage{MnSymbol}

\usepackage{orcidlink}

\renewcommand{\k}{{\ensuremath{\bf{k}}}}
\newcommand{\x}{{\ensuremath{\bf{x}}}}

\newcommand{\refsubfig}[1]{{\color{blue}$#1$)}}

\usepackage{pdfpages}
\usepackage{pgffor}

\makeatletter
\AtBeginDocument{\let\LS@rot\@undefined}
\makeatother

\begin{document}

\title{Dynamical Signatures of Symmetry Broken and Liquid Phases\\ in an $S=1/2$ Heisenberg Antiferromagnet on the Triangular Lattice}
\author{Markus Drescher \orcidlink{0000-0001-9231-1882}}
\thanks{markus.drescher@tum.de}
\affiliation{Department of Physics, Technische Universität München, 85748 Garching, Germany}%

 
\author{Laurens Vanderstraeten \orcidlink{0000-0002-3227-9822}}
\affiliation{Center for Nonlinear Phenomena and Complex Systems, Université Libre de Bruxelles, 1050 Brussels, Belgium}%

\author{Roderich Moessner}
\affiliation{Max-Planck-Institut für Physik komplexer Systeme, 01187 Dresden, Germany}%

\author{Frank Pollmann \orcidlink{0000-0003-0320-9304}}
\affiliation{Department of Physics, Technische Universität München, 85748 Garching, Germany}%
\affiliation{Munich Center for Quantum Science and Technology (MCQST), 80799 Munich, Germany}%

\date{\today}

\begin{abstract}
We present the dynamical spin structure factor of the antiferromagnetic spin-$\frac{1}{2}$ $J_1-J_2$ Heisenberg model on a triangular lattice obtained from large-scale matrix-product state simulations. The high frustration due to the combination of antiferromagnetic nearest- and next-nearest-neighbor interactions yields a rich phase diagram. We resolve the low-energy excitations both in the $120^{\circ}$ ordered phase and in the putative spin-liquid phase at $J_2/J_1 = 0.125$. 
In the ordered phase, we observe an avoided decay of the lowest magnon branch, demonstrating the robustness of this phenomenon in the presence of gapless excitations.
Our findings in the spin-liquid phase chime with the field-theoretical predictions for a gapless Dirac spin liquid, in particular the picture of  low-lying monopole excitations at the corners of the Brillouin zone. We comment on possible practical difficulties of distinguishing proximate liquid and solid phases based on the dynamical structure factor.
\end{abstract}

\maketitle

\paragraph{Introduction}%
Quantum spin liquids (QSLs)~\cite{Wen2002, Balents2010, Savary2016, Knolle2019} are exotic, entangled phases of matter characterized by a lack of magnetic order at zero temperature and the emergence of fractionalized quasiparticle excitations. They have received substantial attention both in theory and experiment, giving rise to various proposals regarding the nature of the candidate QSL phases on frustrated spin systems~\cite{Kitaev2006, Knolle2015, Kitagawa2018, Takagi2019, Sachdev1992, Hastings2000, Ran2007, Iqbal2015, He2017, Liao2017, Fujihala2020, Hu2015, Iqbal2016, Shen2016, Hu2019, Becca2019, Scheie2021, Jiang2022, Tang2022, Xu2023}.
Using state-of-the-art numerics, we study the paradigmatic antiferromagnetic $J_1$-$J_2$ Heisenberg Hamiltonian on a triangular lattice (TLHAF)
\begin{equation}
H = J_1 \sum_{\langle i,j \rangle} {\bf \hat S}_i \cdot {\bf \hat S}_j + J_2 \sum_{\llangle i,j \rrangle} {\bf \hat S}_i \cdot {\bf \hat S}_j\,,
\label{equ:H}
\end{equation}
where $\langle i,j\rangle$ and 
$\llangle i,j \rrangle$ denote pairs of nearest-neighbor and next-nearest-neighbor sites, respectively, 
thereby aiming to find dynamical fingerprints of the distinct phases that have been proposed theoretically.


\begin{figure}[t]
\includegraphics[scale=1]{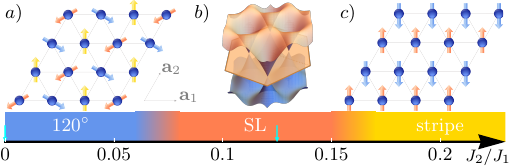}
\caption{Phase diagram of the $J_1-J_2$ antiferromagnetic Heisenberg model on a triangular lattice. At the Heisenberg point $J_2 = 0$, the ground state exhibits a $120^{\circ}$ order ($a$). Around $\frac{J_2}{J_1}\approx 0.07$, there is a transition into a candidate quantum spin-liquid state ($b$). In the illustration, we show the dispersion of the spinons at half-filling in the mean-field solution of the Dirac spin liquid discussed in Sec.~F in the Supplemental Material \cite{supp}.
For larger next-nearest-neighbor couplings, a stripe-ordered phase emerges ($c$). The blue arrows at $J_2 = 0$ and $\frac{J_2}{J_1} = 0.125$ denote the points in the different regimes where we compute the spectral function.}
\label{fig:phase_diagram}
\end{figure}

Following Anderson's original proposal~\cite{Anderson1973} that the ground state of the nearest-neighbor TLHAF could stabilize a resonating valence bond state, there have been intense investigations into the nature of quantum spin models on the frustrated geometry of the triangular lattice
\cite{Zhu2015, Hu2015, Iqbal2016, Saadatmand2016}. Even though the ground state for the nearest-neighbor model has been established to have a coplanar $120^{\circ}$ Néel order~\cite{Bernu1994, Capriotti1999, Chernyshev2006, Starykh2006, Chernyshev2009}, the underlying geometry still provides one of the simplest cases for the emergence of a QSL phase~\cite{Zhu2018}. Adding a next-nearest-neighbor coupling $J_2$, there is classically a phase transition at $\frac{J_2}{J_1} = \frac{1}{8}$ between the $120^{\circ}$ Néel order and a four-sublattice ordered phase with a residual degeneracy~\cite{Jolicoeur1990, Chubukov1992, Zhu2018}. For the quantum model, however, numerical simulations indicate a QSL phase around the point of the classical phase transition for $0.07 \lesssim \frac{J_2}{J_1} \lesssim 0.15$, the nature of which has been under debate~\cite{Zhu2015, Hu2015, Kaneko2014, Hu2019, Iqbal2016, Becca2019, Gong2019, Jiang2022, Tang2022, Shen2016}.
It is followed by a collinear stripe-ordered phase for larger $J_2$~\cite{Zhu2015, Hu2015, Zhu2018} (cf. Fig.~\ref{fig:phase_diagram}).
Despite recent progress in the quest for candidate materials~\cite{Shen2016, Shen2018, Ruan2021}, most promisingly with respect to rare-earth delafossites~\cite{Bordelon2019, Sarkar2019, Dai2021, Scheie2021}, the unambiguous detection of a QSL remains an open issue, with only very recent encouraging reports \cite{Xu2023}.

From a theoretical perspective, the computation of spectral functions of two-dimensional quantum magnets has been a challenge as well, triggering work in 
analytical techniques~\cite{Starykh2006, Chernyshev2006, Chernyshev2009, Mourigal2013, Zheng2006, Zheng2006PRL, Ghioldi2018, Ghioldi2022},
variational Monte Carlo simulations~\cite{Ferrari2018, Becca2019}, and tensor-network approaches~\cite{Gohlke2017, Verresen2018, Chi2022, Kadow2022}.
Using large-scale matrix-product state (MPS) methods \cite{Schollwoeck2011, Hauschild2018, Zaletel2015, Paeckel2019, Vanderstraeten2019}, we compute the dynamical correlations of the system both in the ordered $120^{\circ}$ phase for $J_2=0$ and the adjacent candidate QSL phase at $\frac{J_2}{J_1}=0.125$, 
which allows us to compute the spectral function of the model.  We complement these time-evolution calculations by applying the quasiparticle ansatz, a variational approach for targeting excited states on top of a given MPS ground state for the infinite cylinder directly~\cite{Haegeman2012, VanDamme2021}.
Our central insights are the following. 
First, we find that the  magnon mode in the ordered phase of the {\it isotropic} model is stable even when kinematically its decay is allowed. The mechanism, termed avoided quasiparticle decay, had been reported for a TLHAF with an easy-axis anisotropy~\cite{Verresen2019} but its stability in the presence of gapless excitations had remained in question.
Second, we investigate the dynamical features of the spectral function in the candidate QSL, and compare these to those of the ordered phase. The outstanding agreement between the time evolution and the quasiparticle ansatz suggests that both can be used as complementary methods.
In the candidate QSL phase, we find prominent low-energy excitations at the corners of the Brillouin zone, in agreement with the field-theoretical prediction of triplet monopole excitations in a $U(1)$ Dirac spin liquid (DSL)~\cite{Hermele2005, Song2019, Song2020},
suggesting important gauge fluctuations within the parton construction of the DSL.

\begin{figure}
\includegraphics[scale=1]{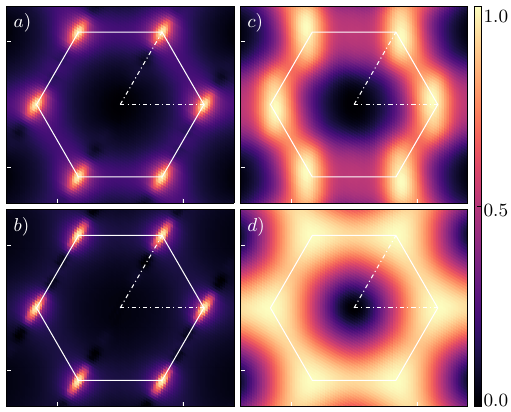}
\caption{Comparison of the static spin structure factor $\chi(\k)$ on a cylinder with $L_y = 6$ from numerical DMRG results (top row) and analytical calculations (bottom row). $a$) DMRG result for the $120^{\circ}$ ordered phase at $J_2=0$ and $b$) the corresponding linear spin-wave theory. A cubic interpolation scheme along the circumference has been used for plotting. The right column shows the static structure factor in the QSL phase at $\frac{J_2}{J_1}=0.125$ for $c$) DMRG and $d$) the analytic result from the Dirac spin-liquid parton theory. The cylinder geometry used is $6 \times 51$ for panels $a$)-$c$) and $6\times 50$ for $d$).}
\label{fig:ssf}
\end{figure}

\paragraph{Numerical Methods}%
We use large-scale MPS simulations to find the ground state and the excitation spectrum of the $J_1-J_2$ TLHAF in two different regimes indicated in  Fig.~\ref{fig:phase_diagram}. The triangular lattice is wrapped onto a cylindrical geometry with periodic boundary conditions along the circumference $L_y$~\cite{Stoudenmire2012, Gohlke2017}.
We consider a YC6-0 geometry~\cite{Zhu2015, Hu2019}, where $L_y=6$ specifies the circumference of the cylinder and $n=0$ determines the boundary condition by identifying sites ${\bf r}$ and ${\bf r} + L_y \,{\bf a}_2 - n \,{\bf a}_1$ with ${\bf a}_{1/2}$ being the primitive vectors of the Bravais lattice [Fig.~\ref{fig:phase_diagram}{\refsubfig{a}}].
We find an MPS ground-state approximation using either the infinite DMRG~\cite{White1992, White1993, McCulloch2008} or the VUMPS algorithm~\cite{ZaunerStauber2018, Vanderstraeten2019}. From this infinite ground-state MPS, the static spin structure factor
\begin{equation}
\chi({\bf k}) = \frac{1}{N} \sum_{i,j} 
e^{-\ui {\bf k} \cdot ({\bf r}_i - {\bf r}_j)}  
\braket{{\bf \hat S}_i \cdot {\bf \hat S}_j}\,,
\label{equ:ssf}
\end{equation}
can be readily obtained as shown in Fig.~\ref{fig:ssf}.




\begin{figure*}
\includegraphics[scale=1]{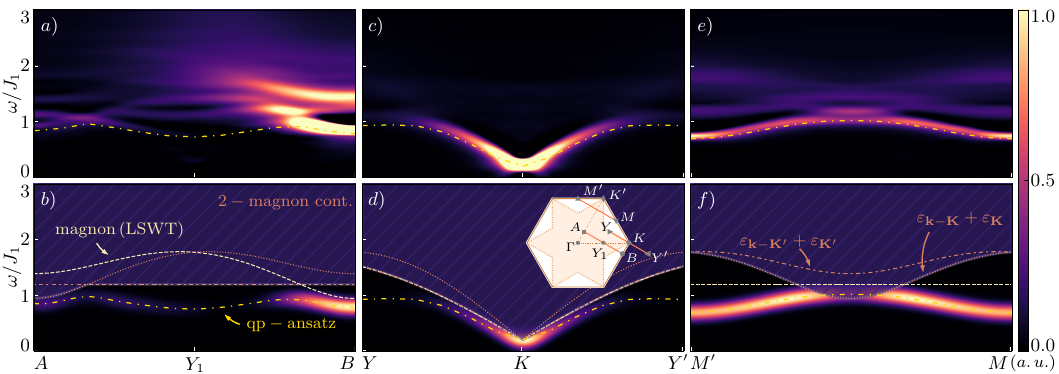}
\caption{Upper row: The spectral function in the ordered phase with only nearest-neighbor coupling ($J_2=0$) along different paths in the Brillouin zone as indicated in the inset in $d$). The golden dot-dashed line shows the result for the lowest-excitation energy obtained from the quasiparticle ansatz. 
Lower row: Results from linear spin-wave theory (LSWT) to be compared to the MPS data in the panels above. The dashed bright line denotes the single-magnon branch. The red dotted line shows the minimum of the two-magnon continuum at ${\bf q} = {\bf K}$, whereas the red dot-dashed line corresponds to the local minimum for ${\bf q} = {\bf K}^{\prime}$. The resulting two-magnon continuum is given by the hatched area.
We indicate again the lowest-energy mode obtained from the quasiparticle ansatz and plot it additionally with a Gaussian broadening of $\sigma = 0.1$ and weighted with the corresponding spectral weights (cf. Sec.~B in the Supplemental Material~\cite{supp}). Note that the finite-size gap determined at ${K}$ from the quasiparticle ansatz has been taken as an energy offset for the linear spin-wave results.
We used system sizes $6\times 126$ and bond dimensions up to $\chi=1500$.}
\label{fig:panel1}
\end{figure*}

We compute the dynamical spin structure factor---or spectral function---from the time-resolved spatial spin-spin correlations of the system
\begin{equation}
S^{+-}({\bf k}, \omega) = \hspace{-0.3em}\int \hspace{-0.3em} \mathrm{d}t  \sum_{j} e^{\ui\omega t - \ui{\bf k} \cdot ({\bf r}_j - {\bf r}_{j_c})} \braket{{\hat S}^{+}_j(t){\hat S}^{-}_{j_c}(0)}\,,
\label{equ:dsf}
\end{equation}
where $j_c$ denotes the center site of the lattice and $\hat S^{\pm}_{j} = \hat S^x_{j} \pm \ui \hat S^y_{j}$ are the spin ladder operators.
%
The simulations are performed on long cylinders of dimension $L_y \times L_x$.
Depending on the parameter regime, we have used $L_x = 51$ or $L_x = 126$.
The dynamical correlations are obtained by applying a local operator $\hat S^-_{j_c}$ at the center of the system and time-evolving the perturbed ground state using the MPO $W_{\text{II}}$ time-evolution algorithm for MPS with long-range interactions~\cite{Zaletel2015, Paeckel2019}. For a time step size of $\delta t = 0.04 \,J_1$, we measured the correlations $C_{i,j_c}^{+-}(t) \equiv \braket{\hat S_i^+(t) \hat S_{j_c}^-}$ every $N_{\mathrm{steps}} = 5$ time steps. We obtain the spectral function by taking the momentum superposition of $C^{+-}_{i,j_c}(t)$, and transforming to frequency space by a numerical integration with a Gaussian envelope and linear-prediction techniques~\cite{White2008}.
The Gaussian window function thereby ensures that the actual simulation data has a large weight whereas the predicted data lie in its tail with the main purpose of suppressing Gibb's oscillations in the Fourier signal (see Sec.~A in the Supplemental Material~\cite{supp} for further details).
The entanglement entropy in the state is expected to grow under unitary time evolution, increasing the difficulty to faithfully represent the quantum state. Hence, the total time that we can access in our simulations is limited by the bond dimension of the MPS. We have used bond dimensions up to $\chi = 1500$ in the ordered and $\chi = 2000$ in the QSL phase with a total simulation time up to $60\,J_1$ and have made sure that the system is large enough in order to avoid boundary artifacts.
Moreover, we have applied operators with defined $k_y$ momentum instead of single-site operators to better resolve the gapless $K$ points (see also Sec.~A in the Supplemental Material \cite{supp}).

We next compare these results to those obtained by  the variational quasiparticle ansatz~\cite{Haegeman2012, VanDamme2021}. Compared to the time evolution, this variational approach involves a numerically cheaper calculation, and provides a good approximation to the energy of the lowest-lying excited state at a certain momentum $\k$. 
As such, it provides the dispersion of isolated modes in the spectrum or the lower edges of continua. In addition, using the variational wavefunctions for the lowest-lying excitations, we can compute the spectral weights and obtain their contribution to the spectral function in the Lehmann representation~\cite{Bera2017}.

\paragraph{$120^\circ$ Ordered Phase}%
The static structure factor $\chi({\bf k})$ exhibits characteristics of the $120^{\circ}$ coplanar Néel order of the ground state at the Heisenberg point $J_2 = 0$ via the well-pronounced maxima at the corners of the Brillouin zone $K$, $K^{\prime}$ and related points
shown in Fig.~\ref{fig:ssf}. Comparing with the results from linear spin-wave theory (LSWT), where the classical three-sublattice $120^{\circ}$ order is weakened by quantum fluctuations (see Sec.~E in the Supplemental Material \cite{supp}
for details of the calculation), we find conclusive agreement.

Figure~\ref{fig:panel1} shows the results for the spectral function from the MPS simulations for $J_2 = 0$. Previous investigations of the low-energy excitations of the ordered phase beyond LSWT using semiclassical approaches such as higher-order spin-wave theory~\cite{Capriotti1999, Starykh2006, Chernyshev2009} or series expansions~\cite{Zheng2006, Zheng2006PRL} already obtained strong renormalizations of the magnon dispersion mainly caused by interactions between the single quasiparticle branches and the magnon continuum.
One prominent feature is a pronounced rotonlike minimum at the $M$ point and symmetry-related points (the centers of the edges of the Brillouin zone). 
Our numerics confirms this characteristic as shown in Fig.~\ref{fig:panel1}\refsubfig{e}.
The spectral weight is concentrated at the $M$ points in a smaller frequency window, leading to a distinct maximum in the intensity although the integrated spectral weight of the lowest branch exhibits a local minimum (cf. Sec.~C in the Supplemental Material \cite{supp}).
Using the quasiparticle ansatz, we find good agreement for the overall spectral weight of the lowest renormalized magnon branch.
The energy dispersion from this ansatz with the associated spectral weight displayed as a Gaussian broadening of $\sigma = 0.1$ is plotted in Fig.~\ref{fig:panel1}\refsubfig{f}. 
The bright dashed line denotes the magnon branch from LSWT. We observe a strong downward renormalization that is consistent with previous findings~\cite{Chernyshev2009, Mourigal2013}.
The lower bound of the two-magnon continuum is given by the minimum of the decays $\varepsilon_{\bf k} = \varepsilon_{{\bf k}- {\bf q}} + \varepsilon_{{\bf q}}$ with ${\bf q} = {\bf K}$ and ${\bf q} = {\bf K^{\prime}}$~\cite{Chernyshev2009}.
Both dispersions are schematically shown in Fig.~\ref{fig:panel1}\refsubfig{f}.
In our simulations on an $L_y=6$ cylinder, the spectral weight distribution above the isolated lowest mode in Fig.~\ref{fig:panel1}\refsubfig{e} exhibits features of both decay channels and the non-interacting magnon line alongside some distortions around the touching points. It remains an open question if this is a finite-size effect or whether it survives in the thermodynamic limit.

\begin{figure*}
\includegraphics[width=\linewidth]{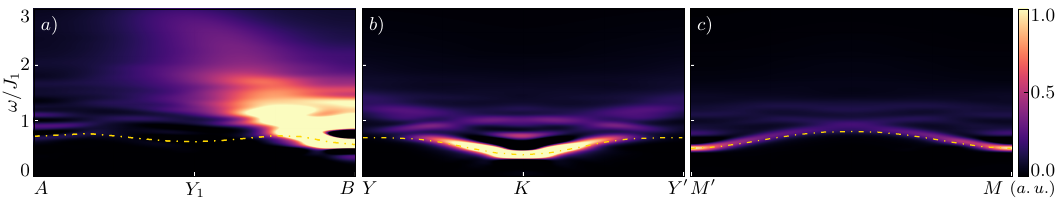}
\caption{Spectral function in the candidate QSL phase at
$\frac{J_2}{J_1} = 0.125$ along the same momentum cuts as in Fig.~\ref{fig:panel1} obtained on a cylinder geometry with $L_y=6$ and $L_x=51$. 
The dot-dashed line denotes the corresponding lowest-energy mode from the quasiparticle ansatz. We used bond dimension $\chi = 2000$ for these simulations.
}
\label{fig:panel2}
\end{figure*}

A rotonlike minimum, however, does not only occur at the $M$ points, but also at $Y_1$, the midpoint between $A$ and $B$ [cf. inset Fig.~\ref{fig:panel1}\refsubfig{d}]. This was first suggested by Verresen et al.~\cite{Verresen2019} in the case of an antiferromagnetic triangular Heisenberg model with a small anisotropic term that makes it numerically more tractable by slightly gapping out the Goldstone modes.
Moreover, also variational Monte Carlo simulations~\cite{Becca2019} have found a similar feature that can be associated with avoided quasiparticle decay due to strong interactions between the lowest mode and the two-magnon continuum~\cite{Verresen2019}. 
Fig.~\ref{fig:panel1}\refsubfig{a} shows the spectral function of the pure Heisenberg model on the triangular lattice along the cut $A-B$. 
Besides a pronounced maximum at $B$ and a diffuse continuum reaching up to energies of roughly $3J_1$, we again observe remnants of the single-magnon mode and the decay channels representing the minima of the energy surface described by the lower boundary of the two-magnon continuum.
The lowest-energy mode, however, stays beneath the onset of the continuum, supporting the previous conjecture that avoided magnon-decay is a valid feature of the theory~\cite{Verresen2019, Becca2019}.
The variation of the integrated spectral weight of the repelled magnon mode is confirmed within the quasiparticle ansatz
(see Sec.~C in the Supplemental Material \cite{supp}).

The Goldstone modes at the corners of the Brillouin zone ${K}$ and ${K^{\prime}}$ exhibit a high concentration of spectral weight [Fig.~\ref{fig:panel1}\refsubfig{c}]. This agrees with previous numerical and experimental findings~\cite{Becca2019, Macdougal2020, Scheie2021}. 
Due to finite-size effects, the magnon modes develop a gap.
The corresponding gap determined at ${K}$ for the quasiparticle ansatz has been taken as a reference offset when plotting the analytic linear spin-wave results in Fig.~\ref{fig:panel1}.


\paragraph{Stripe-Ordered Phase}%
For $\frac{J_2}{J_1}\gtrsim 0.15$, we obtain a stripe-ordered phase. The DMRG algorithm applied here for the ground-state optimization chooses a certain symmetry-broken state as the ground state.
As a result, the static structure factor displays a peak at $M^\prime$, but not at $M$.
In accordance with analytical expectations, we observe a gapless Goldstone mode at $M^\prime$. 
Plots of the spectral function for $\frac{J_2}{J_1}=0.55$ deep in the symmetry-broken phase are provided in Sec.~D in the Supplemental Material~\cite{supp}.
We observe clear magnon modes in good agreement with the predictions from linear spin-wave theory.

From pure LSWT, however, we expect accidental zero modes at the edge centers of the Brillouin zone where no Goldstone mode resides. 
In contrast, our results suggest a clear finite gap that exceeds the finite-size gap of the system (cf. Sec.~D in the Supplemental Material~\cite{supp} for details). 
This is in line with the established understanding that quantum fluctuations in the Heisenberg model gap out the accidental zero modes \cite{Jolicoeur1990, Chubukov1992, Willsher2023}.

\paragraph{Candidate QSL Phase}%
While the existence of a QSL phase around the classical phase transition point $\frac{J_2}{J_1} = 0.125$ is rather well-established, the precise nature of this phase is  highly debated so far, reaching from gapped $Z_2$ spin liquids to gapless $U(1)$ Dirac spin liquids (DSL) or chiral spin liquids~\cite{Zhu2015, Iqbal2016, Becca2019}.
%
Note that the $J_1-J_2$ triangular Heisenberg model on an even cylinder comprises two different topological sectors in the candidate QSL phase~\cite{Zhu2015, Hu2015}. The sector of the isotropic ground state can be reached by adiabatically inserting a flux $\theta = 2\pi$~\cite{Hu2019} (cf. Sec.~A in the Supplemental Material \cite{supp} for further details). We focus on this sector for our simulations and all subsequent results.
%

%
Figure~\ref{fig:panel2} shows the dynamical structure factor for $\frac{J_2}{J_1} = 0.125$.
%
We observe a softening of the minima at the $M$ points compared to the ordered phase, which can be attributed to the existence of spinon bilinears at the centers of the edges~\cite{Song2019} in a $U(1)$ DSL.
Apart from this, the continuum is shifted downwards with varying strength between $M^{\prime}$ and $M$, diminishing the distance in energy between the lowest-energy mode and the onset of the continuum. This is in accordance with variational Monte Carlo data that suggests a vanishing separation of the continuum from lowest-energy excitations in the QSL phase~\cite{Becca2019}.
Even though we expect the bilinear excitations to become gapless in the two-dimensional limit (i.e., $L_y \rightarrow \infty$), we observe a clear gap in the spectral function. This can be related to the geometry of the $L_y=6$ cylinder: The accessible momenta do not include the spinon Dirac cones at $\pm {\bf Q} = \pm \left(\frac{\pi}{2}, \frac{\pi}{2\sqrt{3}}\right)$~\cite{Ferrari2021} (see Sec.~G in the Supplemental Material \cite{supp}),
which gaps out the corresponding spectral function.
The spectral maximum at $B$ survives across the transition from the ordered phase to the QSL phase [Fig.~\ref{fig:panel2}\refsubfig{a}] as well as the feature of the lowest mode with almost vanishing weight being repelled from the continuum at intermediate energies [Fig.~\ref{fig:panel2}\refsubfig{a}]\nobreakdash---although the spectral function shows different distinct modes below the more pronounced continuum. 

Perhaps the most striking aspect, however, is the structure of the excitations at the $K$ points. As for the Goldstone mode in the symmetry-broken phase, there is a minimum in the candidate QSL phase, albeit with a flatter dispersion and a rich structure of the distribution of the spectral weight above the minimum in contrast to the $120^{\circ}$ phase.
The locations of the minima at ${K}$ and ${K^{\prime}}$ and related points are in accordance with the field-theoretical predictions by Song \textit{et al.}~\cite{Song2019, Song2020}:
They report the occurrence of triplet monopole excitations at the corners of the Brillouin zone for a $U(1)$ DSL on a triangular lattice. The comparison of Fig.~\ref{fig:panel2}\refsubfig{b} with variational Monte Carlo data for a DSL ansatz \cite{Becca2019, Ferrari2020} supports this conjecture.
At the transition to the ordered phase for smaller $J_2$ coupling, the monopole operators whose quantum numbers correspond to the $K$ points condense~\cite{Song2019, Dupuis2019}, thus building up the familiar three-sublattice order. The comparison of the static structure factor sustains this theory: 
Although in the QSL phase at $\frac{J_2}{J_1}=0.125$, $\chi({\bf k})$ still exhibits maxima in the intensity around the $K$ points of the Brillouin zone (Fig.~\ref{fig:ssf}), the structure factor obtained from DMRG simulations has a much broader and more diffuse structure than in the ordered phase at $J_2 = 0$, which compares well with the analytic result for a $U(1)$ DSL [Fig.~\ref{fig:ssf}\refsubfig{d}], suggesting qualitative agreement notwithstanding the finite-cylinder effects.

\paragraph{Conclusions}%
We have studied the dynamical properties of  ordered and candidate spin-liquid phases in the triangular lattice. We find excellent agreement 
between the time evolution and the quasiparticle ansatz. In the ordered phase, we  observe the avoided decay of the lowest magnon branch previously shown to occur in an anisotropic Heisenberg model \cite{Verresen2019}.
In the candidate QSL phase, our numerical MPS results for the Heisenberg model on a cylinder show good agreement with the results obtained from variational Monte Carlo simulations~\cite{Becca2019}, where the gapless $U(1)$ Dirac spin-liquid nature of the ground state is assumed.
Given that the method we apply does not depend on any previous knowledge or assumptions on the wave function of the ground state, this validates the variational $U(1)$ DSL approach.

In closing, we would like to make the following observation. While the responses of the two phases can be visually quite distinct, see, e.g., the broad features in the QSL in Fig.~\ref{fig:ssf}, the differences are perhaps overall less striking than one might have hoped for. Indeed, inspecting Figs.~\ref{fig:panel1} and \ref{fig:panel2} yields copious similarities between the two dynamical structure factors. This underlines the challenge inherent in discriminating proximate phases whose universal low-energy physics may be fundamentally distinct, but which may nonetheless harbor similar finite-time and short-distance correlations. Of course, if a reliably determined microscopic Hamiltonian for a material under consideration is available, our microscopic approach can be used for a direct {\it quantitative} validation of the correspondence between experimental results and theoretical interpretation. Absent this, it may very well be possible to account for salient finite-time and short-distance correlations from different starting points. That this should be the case of course is what makes the universal response so special by contrast; but in practice, this can pose a formidable challenge in the interpretation of experimental data.

\paragraph{Note added}%
While finalizing the current draft, we became aware of a similar work about the spectral function of different phases on the TLHAF \cite{Sherman2022}.

\paragraph{Acknowledgments}%
We would like to thank Ruben Verresen, Wilhelm Kadow and Johannes Knolle for helpful discussions and Francesco Ferrari and Federico Becca for providing additional data for spectral functions from variational Monte Carlo simulations.
F.P. acknowledges support of the European Research Council
(ERC) under the European Union's Horizon 2020 research
and innovation program (Grant No. 771537).
F.P. also acknowledges the support of the Deutsche
Forschungsgemeinschaft (DFG, German Research Foundation) under Germany’s Excellence Strategy Grant No. EXC-2111-
390814868. F.P.’s research is part of the Munich Quantum Valley, which is supported by the Bavarian state
government with funds from the Hightech Agenda Bayern Plus. This work was in part supported by the Deutsche
Forschungsgemeinschaft under Grant No. SFB 1143 (Project No. 247310070) and cluster of excellence 
ct.qmat (EXC 2147, Project No. 390858490). L.V. is supported by the Research Foundation Flanders (FWO) via Grant No. FWO20/PDS/115.

\bibliographystyle{apsrev4-2}
\bibliography{heisenberg_triangular_v3_arXiv}



\section*{Supplementary material}

# Supplemental Material: Dynamical Signatures of Symmetry Broken and Liquid Phases in an $S = 1/2$ Heisenberg Antiferromagnet on the Triangular Lattice

Markus Drescher,[1, *] Laurens Vanderstraeten,[2] Roderich Moessner,[3] and Frank Pollmann[1, 4]

[1]*Department of Physics, Technische Universität München, 85748 Garching, Germany*
[2]*Center for Nonlinear Phenomena and Complex Systems,*
*Université Libre de Bruxelles, 1050 Brussels, Belgium*
[3]*Max-Planck-Institut für Physik komplexer Systeme, 01187 Dresden, Germany*
[4]*Munich Center for Quantum Science and Technology (MCQST), 80799 Munich, Germany*
(Dated: November 14, 2023)

## A: Spectral Functions from Dynamical Correlations

We compute the dynamical spin structure factor from time-dependent spin correlations according to equation (3) of the main text. The quantum state is represented by an MPS wrapped around a cylinder with circumference $L_y = 6$. The ground state is via an iDMRG simulation [1] on a $L_y \times 3$ unit cell. By periodically repeating the unit cell, we can build a large system whose center site $j_c$ is perturbed by a local $\hat{S}^-$ operator. In the antiferromagnetically ordered phase at $J_2 = 0$, we used cylinders with $L_y = 6$ and $L_x = 51$ as well as $L_x = 126$, whereas in the QSL phase at $\frac{J_2}{J_1} = 0.125$, the system size was given by $L_y = 6$ and $L_x = 51$.

The MPO $W_{II}$ time-evolution algorithm [2, 3] is suited to fully capture long-range interactions as they occur in the two-dimensional system mapped onto a cylinder. For the final large-scale tensor-network simulations, we terminated the iMPS of the ground state for the enlarged system by applying the dominant left and right eigenvector of the associated transfer matrix to the system which have been found using a power method [4]. The obtained finite MPS can be optimized at the boundaries using DMRG on a segment of a few of the outmost rings. Thus we can run the simulations on a large finite MPS with open boundary conditions in $x$-direction. This results in an effective speed-up of the time-evolution simulation compared to the infinite system. The bulk physics does not change.

The time-step size we used for the data presented is $\delta t = 0.04 J_1$. The correlations are computed every $N_{\text{steps}} = 5$ simulation steps. We apply $U(1)$ charge conservation during the time evolution (i.e., the total $S^z$ quantum number is conserved) [5–7]. The protocol how to compute the spin structure factor from the dynamical correlations involves several single steps: First, we can interpolate the correlation data for times between $t$ and $t + N_{\text{steps}} \cdot \delta t$. We apply then a discrete Fourier transformation in space that yields the momentum-resolved time-dependent data. For each point in momentum space, we perform a linear prediction that extends the time series by $f_{\text{LinPre}} \cdot T_{\text{sim}}$, where $T_{\text{sim}}$ is the total simulation time. The choice of $f_{\text{LinPre}} = 10$ has turned out to be suitable for the systems under consideration, where always $m_{\text{LinPre}} = 20$ values of the time series have been taken into account to predict the next value. To conclude the computation, we performed a Fourier transformation of the time series convoluted with a Gaussian window function to prevent Gibb's oscillations due to the limited reachable simulation time. The Gaussian distribution was chosen to have a defined value of $\alpha$ at $t = T_{\text{sim}}$, i.e., the broadening is given as

$$\sigma = \sqrt{\frac{T_{\text{sim}}^2}{-2 \ln(\alpha)}}. \tag{1}$$

For the plots shown in the main text and the Supplemental Material, we chose $\alpha = 0.01$, corresponding to a Gaussian broadening of $\sigma \approx 19.8 J_1$ for $T_{\text{sim}} = 60 J_1$. As a result of these parameter settings, the convolution with the Gaussian function distributes the main weight to the actual simulation data whereas the linearly predicted data occurs only in the tail of the Gaussian envelope, thus avoiding a sharp cutoff of the time series and consequently Gibb's oscillations in the Fourier transform.

*$k_y$ Resolved Time Evolution.*—The method above provides the dynamical correlation data for all available cuts through the full two-dimensional Brillouin zone that are accessible on an $L_y \times L_x$-geometry (cf. Fig. 10). Since the entanglement entropy grows under time evolution of the perturbed ground state, the bond dimension $\chi$ restricts the feasible simulation times and determines the accuracy of the results. Consequently, if one separates the contributions from the different cuts in the Brillouin zone, the increasing entanglement for the corresponding momentum quantum numbers can be captured more faithfully and the overall accuracy for a given bond dimension is expected to improve.

To achieve this, we can apply a local operator with well-defined momentum $k_y(m_2) = 2\pi \frac{m_2}{L_y}$ ($m_2 \in \mathbb{Z}$) in $\mathbf{b}_2$-direction instead of a single-site operator:

$$\hat{S}_{n_1}^-(k_y) = \frac{1}{\sqrt{L_y}} \sum_{j=0}^{L_y - 1} e^{\mathbf{j} \cdot k_y} \hat{S}_{n_1 \cdot \mathbf{a}_1 + j \cdot \mathbf{a}_2}^-. \tag{2}$$

The primitive lattice vectors are $\mathbf{a}_1 = (1, 0)^T$ and $\mathbf{a}_2 = \left(\frac{1}{2}, \frac{\sqrt{3}}{2}\right)^T$ with the corresponding reciprocal lattice vec-

tors $\mathbf{b}_1 = \left(2\pi, -\frac{2\pi}{\sqrt{3}}\right)^T$ and $\mathbf{b}_2 = \left(0, \frac{4\pi}{\sqrt{3}}\right)^T$. The time evolution can be achieved by the same techniques as before. The correlations are taken with respect to the hermitian conjugate operator

$$C_{k_y}(n_1', n_1; t) = \langle e^{\mathrm{i}Ht} \left(\hat{S}_{n_1'}^-(k_y)\right)^\dagger e^{-\mathrm{i}Ht} \hat{S}_{n_1}^-(k_y)\rangle, \quad (3)$$

thus yielding effectively one-dimensional correlations in the $x$-coordinate. Fig. 1 shows the time-evolved bipartite entanglement entropy at the center of the system (upper row) and the dynamical correlation function (bottom row) for a momentum-resolved simulation with $k_y = 2\pi \frac{m_2}{L_y}\big|_{m_2=2}$ (left column) and for the general procedure using a single-site operator (right column) in the symmetry-broken phase at $J_2 = 0$.

As mentioned in the main text, there are two distinct sectors in the QSL phase on a $L_y = 6$ cylinder [8, 9]. The even sector is found directly by the DMRG simulation. In this sector, the entanglement spectrum on the bonds is symmetric around the total $S^z$ quantum number $q_z = 0$. By adiabatically inserting a flux of $\theta = 2\pi$ [4], the wave function transitions into the odd sector, which results in a state whose entanglement spectrum is symmetric around $q_z = \frac{1}{2}$, indicating the existence of a spinon quasiparticle with $S^z = \pm\frac{1}{2}$ on each boundary [9]. The ground state energy in the odd sector is slightly lower and the correlations are more isotropic.

Fig. 2 shows the static structure factor from the DMRG simulation for the 120° ordered phase and the candidate quantum spin-liquid phase at $\frac{J_2}{J_1} = 0.125$. In contrast to Fig. 2 in the main text, we do not apply an interpolation scheme here. The observable features are identical. Due to the small finite cylinder circumference of $L = 6$, the data displays a structure of discrete stripes along the direction of $\mathbf{b}_1$.

Figs. 3 and 4 show the dynamical spin structure factors for $J_2 = 0$ and $\frac{J_2}{J_1} = 0.125$ in a linear and a logarithmic color scale for all avaibable cuts through the Brillouin zone for various bond dimensions. The simulation resolving the gapless Goldstone mode at the $K$ point has turned out to be numerically particularly challenging. Therefore, the best converged data presented has been obtained from a $k_y$ resolved time evolution with $m_2 = 2$. Note that the maximum spectral intensity has been bound by a maximum value to reveal lower-intensity features of the excitation spectrum. The maximum spectral intensity shown in each subplot has been normalized to one. No cutoff has been applied for $J_2 = 0$ along $M' - M$ and $\Gamma - M''$ and in the QSL phase along $M' - M$. To ensure that the data has converged, we ran multiple simulations with different maximal bond dimensions $\chi$, time step sizes $\delta t$ and system sizes $L_x$. $L_x$ and $\chi$ strongly restrict the reliably reachable simulation times. We performed a time evolution up to $T_{\mathrm{sim}} = 60 J_1$ with bond dimensions up to $\chi = 1500$ and $\chi = 2000$ in the case

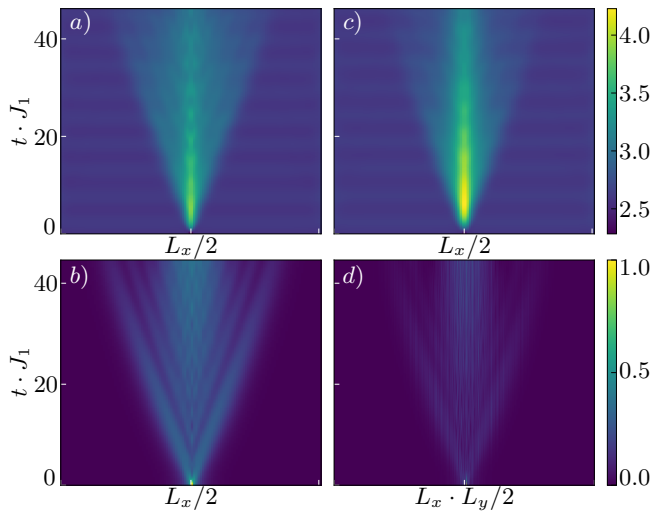

FIG. 1. $a)$ Entanglement cone and $b)$ correlation spreading after perturbing the ground state at $J_2 = 0$ with a momentum-resolved operator versus the corresponding quantities in the right column for a single-site perturbation operator $S_{j_c}^-$. Note that in $d)$, the spatial index runs spirallike through the whole system, not only along the x-direction as in the other cases. The simulations were performed on a $6 \times 126$ cylinder with bond dimensions $\chi = 1300$ and $\chi = 1500$ for the momentum-resolved operator and the single-site operator respectively and time-step size $\delta t/J_1 = 0.04$.

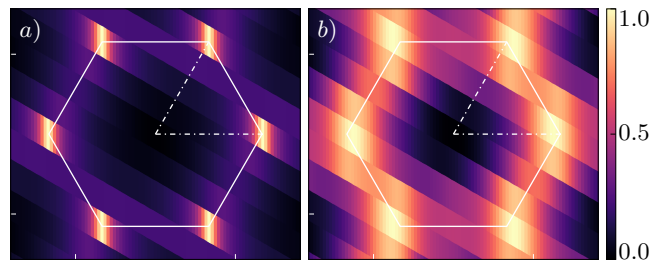

FIG. 2. Static structure factor on a cylinder with dimensions $6 \times 51$ in $a)$ the 120° ordered phase and $b)$ in the candidate spin-liquid phase. We show the same data as in Fig. 2 in the main text, but without using an interpolation scheme along the $\mathbf{a}_2$ direction of the lattice (the circumference of our cylinder).

of $J_2 = 0$ and $\frac{J_2}{J_1} = 0.125$ respectively. From comparing the first two rows in both Fig. 3 and 4, we can evaluate the convergence of the spectral function since the respective simulations have been run for different bond dimensions. Remarkably, Fig. 3$e)$ shows a smoother and hence better converged dispersion with a higher concentration of spectral weight at the $K$-point than Fig. 3$d)$ even though for the latter, we used $\chi = 1500$ and for the former $\chi = 1300$. Fig. 3$e)$, however, was obtained using the $k_y$ momentum-resolved algorithm.

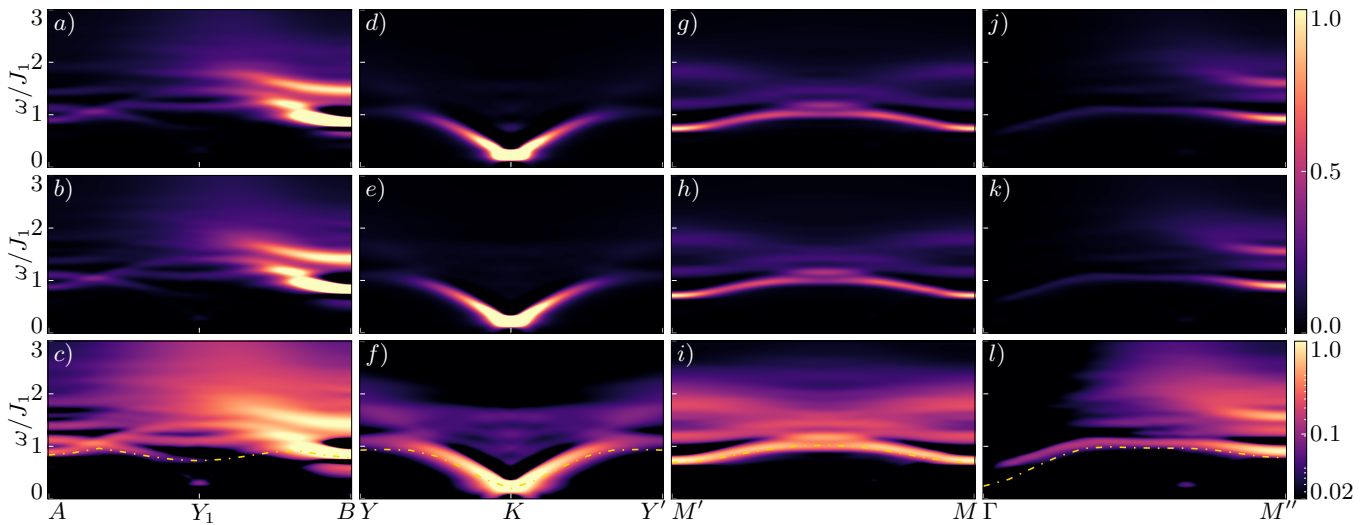

FIG. 3. Convergence of the dynamical spin structure factor $S^{+-}(\mathbf{k}, \omega)$ in the 120° ordered phase at $J_2 = 0$. The simulations have been done on a cylinder of size 6 × 126. We used bond dimensions $\chi = 1200$ [$a$), $g$) and $j$)] and $\chi = 1500$ [$b$), $d$), $h$) and $k$)]. For Fig. $e$), we applied the $k_y$ resolved time-evolution algorithm with $\chi = 1300$. The two upper rows are plotted with a linear color scale. The third row shows the same data as the second, but in a logarithmic scale. The colormap is normalized to the maximum intensity shown in the single subfigure. The golden dot-dashed line denotes the energy of the lowest branch as obtained from the quasiparticle ansatz.

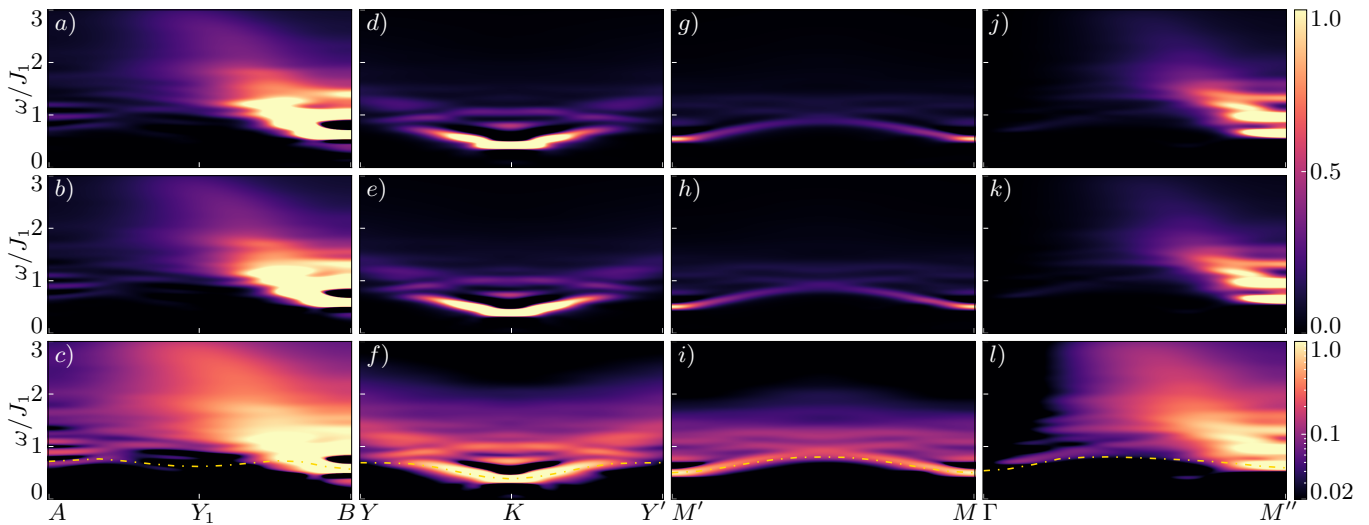

FIG. 4. Convergence of the dynamical spin structure factor $S^{+-}(\mathbf{k}, \omega)$ in the QSL phase at $\frac{J_2}{J_1} = 0.125$. The simulations have been done on a cylinder of size 6 × 51. The first row shows data for a bond dimension of $\chi = 1500$, the second for $\chi = 2000$. The third row displays the same data as the second, but plotted with a logarithmic color scale instead of a linear one. The colormap is normalized to the maximum intensity shown in the single subfigure. As before, the golden dot-dashed line denotes the energy of the lowest branch as obtained from the quasiparticle ansatz.

## B: Quasiparticle Ansatz

In our work, we have also used the MPS quasiparticle ansatz [10, 11], which allows us to target momentum states on top of an infinite MPS ground state directly in the thermodynamic limit. Here we use the adaptation to cylinder geometries where the $k_y$ momentum is used as an approximate quantum number [12]. In ad-

dition, we have used $SU(2)$ quantum numbers in the MPS ground state and quasiparticle ansatz, allowing us to target triplet excitations explicitly. The optimization of the variational energy of a quasiparticle ansatz for given momenta $(k_x, k_y)$ gives us the energy dispersion of the lowest-lying excitations in the system. The variational wavefunction also allows us to compute the spectral weights of these states such that we can compute

their contributions to the spectral function.

Let us start from the definition of the spectral function

$$S(\mathbf{k}, \omega) = \sum_\alpha \int dt e^{i\omega t} \langle \Psi_0 | e^{iHt} (\hat{S}_\mathbf{k}^\alpha)^\dagger e^{-iHt} \hat{S}_\mathbf{k}^\alpha | \Psi_0 \rangle, \quad (4)$$

where we use the sum of the three spin components $\alpha$ such that we do not break $SU(2)$ symmetry. We have defined a momentum operator formally as

$$\hat{S}_\mathbf{k}^\alpha = \frac{1}{\sqrt{L_x L_y}} \sum_\mathbf{r} e^{i\mathbf{k} \cdot \mathbf{r}} \hat{S}_\mathbf{r}^\alpha \quad (5)$$

in the limit $L_x \to \infty$ of an infinitely long cylinder. We can now project the spectral function on the first low-lying excitations, resulting in

$$S_{\text{low}}(\mathbf{k}, \omega) = \sum_\gamma \int dt e^{i\omega t} e^{-i\Delta_\gamma(\mathbf{k}) t} \sum_\alpha \left| \langle \Phi_\mathbf{k}^\gamma | \hat{S}_\mathbf{k}^\alpha | \Psi_0 \rangle \right|^2 \quad (6)$$

$$= \sum_\gamma \sum_\alpha \left| \langle \Phi_\mathbf{k}^\gamma | \hat{S}_\mathbf{k}^\alpha | \Psi_0 \rangle \right|^2 2\pi \delta(\omega - \Delta_\gamma(\mathbf{k})) \quad (7)$$

where $\gamma$ labels the excited states with excitation energy $\Delta_\gamma(\mathbf{k})$. In order to plot this projected spectral function, we convolute the delta signal with a Gaussian broadening

$$2\pi\delta(\omega) \to \frac{\sqrt{2\pi}}{\sigma} \exp\left(-\frac{\omega^2}{2\sigma^2}\right) \quad (8)$$

Note that along the momentum cut $M'$-$M$ through the Brillouin zone [Fig. 3f)], the lowest branch is made up out of two separate but almost degenerate excitations. Therefore, we plot the averaged energy and the added spectral weight [Fig. 5c)].

### C: Integrated Spectral Weight

The energy dispersion for the lowest branch obtained using the two different approaches described in the two preceding sections show good agreement for the accessible momenta in the Brillouin zone (cf. Fig. 3). One specific feature is the repulsion of the magnon branch between $A$ and $B$ in the ordered phase as discussed in the main text. One prominent characteristic of this renormalized branch is the almost vanishing spectral weight near the energetic minimum. To verify the reliability of the two methods, we compare the integrated spectral weight for the lowest branch obtained from the dynamical correlation function and the quasiparticle ansatz.

The latter allows us to access the variational wavefunction for the lowest-lying excitations, thus enabling us to compute the integrated spectral weight of the corresponding excitation modes. From the Lehmann representation of equation (7), we find for the integrated spectral weight of the lowest-energy excitations, denoted by the wavefunctions $|\Phi_\mathbf{k}^\gamma\rangle$:

$$\int d\omega\, S_{\text{low}}(\mathbf{k}, \omega) = 2\pi \sum_\gamma \sum_\alpha \left| \langle \Phi_\mathbf{k}^\gamma | \hat{S}_\mathbf{k}^\alpha | \Psi_0 \rangle \right|^2. \quad (9)$$

Due to the translation invariance of the ground state, the spectral function in equation (3) of the main text is equivalent to

$$S^{+-}(\mathbf{k}, \omega) = \frac{1}{N} \int dt \sum_{i,j} e^{i\omega t - i\mathbf{k} \cdot (\mathbf{r}_i - \mathbf{r}_j)} \langle e^{iHt} \hat{S}_i^+ e^{-iHt} \hat{S}_j^- \rangle, \quad (10)$$

where $N = L_y \cdot L_x$ is the number of sites. In the Lehmann respresentation, this reads

$$S^{+-}(\mathbf{k}, \omega) = \sum_n \left| \langle n | \hat{S}_\mathbf{k}^- | \Psi_0 \rangle \right|^2 2\pi \delta(\omega + E_0 - E_n). \quad (11)$$

$\{|n\rangle\}$ denotes the set of energy eigenstates with energies $E_n$ and $E_0$ is the ground state energy. For an $SU(2)$ invariant state, it holds that

$$\sum_n \left| \langle n | \hat{S}_\mathbf{k}^\pm | \Psi_0 \rangle \right|^2 = \sum_n \left[ \left| \langle n | \hat{S}_\mathbf{k}^x | \Psi_0 \rangle \right|^2 + \left| \langle n | \hat{S}_\mathbf{k}^y | \Psi_0 \rangle \right|^2 \right]. \quad (12)$$

Hence, we can establish the following relation for the integrated spectral weight for an $SU(2)$ symmetric ground state:

$$\sum_n \sum_\alpha \left| \langle \Phi_\mathbf{k}^n | \hat{S}_\mathbf{k}^\alpha | \Psi_0 \rangle \right|^2 = \frac{1}{2\pi} \int d\omega S(\mathbf{k}, \omega) \quad (13)$$

$$= \chi(\mathbf{k}) \quad (14)$$

$$= \frac{3}{2} \sum_n \left| \langle n | \hat{S}_\mathbf{k}^- | \Psi_0 \rangle \right|^2 \quad (15)$$

$$= \frac{3}{4\pi} \int d\omega S^{+-}(\mathbf{k}, \omega). \quad (16)$$

Note that the total integrated spectral weight is identical to the static spin structure factor in equation (2) of the main text. Hence, the spectral weight for the lowest-energy excitations that we obtain from the quasiparticle ansatz is related to the integrated spectral weight from the dynamical structure factor as computed above via

$$\sum_\gamma \sum_\alpha \left| \langle \Phi_\mathbf{k}^\gamma | \hat{S}_\mathbf{k}^\alpha | \Psi_0 \rangle \right|^2 = \frac{3}{4\pi} \int d\omega S_{\text{low}}^{+-}(\mathbf{k}, \omega). \quad (17)$$

Here, we integrate only over the lowest-excitation branch in the spectral function for a given momentum $\mathbf{k}$.

Fig. 5 shows the spectral weights obtained from the two distinct methods for the lowest-energy branch on the different lines through the Brillouin zone as displayed in Fig. 3 of the main text. We observe good agreement between the absolute numbers, which underlines the validity of the results.

### D: Spectral Functions in the Stripe Phase

In Fig. 6, we provide spectral data for $\frac{J_2}{J_1} = 0.55$ in the stripe-ordered phase. As mentioned in the main text, the DMRG algorithm with $U(1)$ charge conservation we use

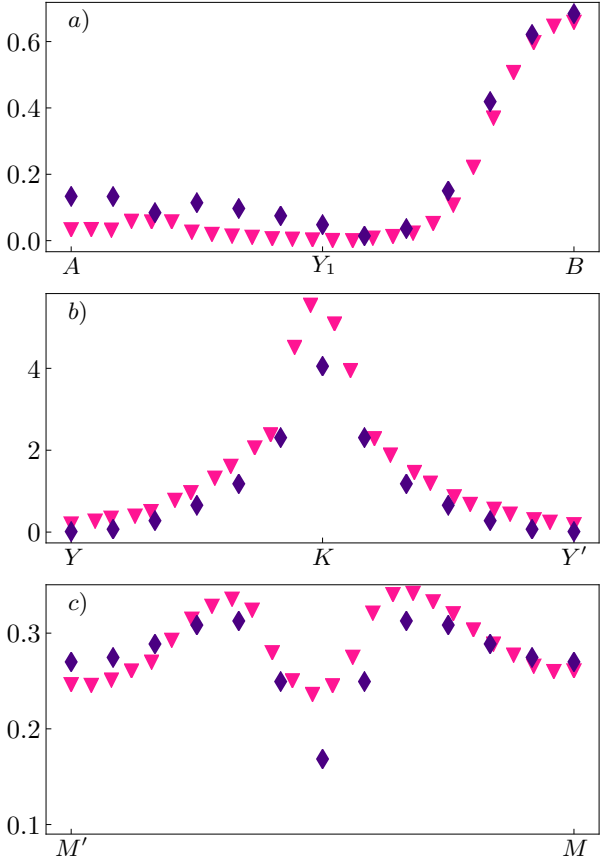

FIG. 5. Comparison of the integrated spectral weight for the lowest-energy excitation $\sum_\gamma \sum_\alpha \left| \langle \Phi_\mathbf{k}^\gamma | \hat{S}_\mathbf{k}^\alpha | \Psi_0 \rangle \right|^2$ (cf. equation (17)) from the quasiparticle ansatz (blue diamonds) and the full spectral function (red triangles) along the different lines in the Brillouin zone in Fig. 3 (main text) for the symmetry broken phase at $J_2 = 0$.

here to find the ground state chooses a certain symmetry-broken state as the ground state. The numerically determined ground state is hence not a superposition of all possible orientations of the stripe-ordered state and its static structure factor exhibits a peak at $M'$, but not at $M$.

We observe that a gapless Goldstone mode develops at $M'$ as expected. Overall, deep in the symmetry-broken phase, one can observe clear magnon modes with well-defined dispersions. They agree well with the results from linear spin-wave theory as shown in Fig. 6. When plotting the linear spin-wave dispersion, we added a finite size gap to the energy as it develops in the system under consideration. The gap has been extracted at the maximum of the spectral weight at momentum $M'$ where the Goldstone mode becomes gapless. We found a numerical value of

$$\Delta E_{\text{finite-size}} \left( \frac{J_2}{J_1} = 0.55 \right) \approx 0.169 \, J_1 . \tag{18}$$

In comparison, the finite size gap as obtained from the quasiparticle ansatz at the expected gapless $K$ point for the Goldstone in the 120° ordered phase at $J_2 = 0$ was found to be

$$\Delta E_{\text{finite-size}} \left( J_2 = 0 \right) \approx 0.191 \, J_1 . \tag{19}$$

From linear spin-wave theory, we expect an accidental zero mode at the edge-centers $M$ and $M''$ where no Goldstone modes are located. In the light of equation (18), however, our results indicate a clear finite gap that exceeds the finite-size gap of the system [see Figs. 6c) and d)]. This is in accordance with the settled insight that quantum fluctuations in Heisenberg models gap out the accidental zero modes [13–15].

### E: Spin-Wave Theory

**Linear Spin-Wave Theory in the 120° Ordered Phase**

We calculated the magnon dispersion for the nearest-neighbor Heisenberg model on the triangular lattice in linear spin-wave theory [16–20] to be compared with our numerical DMRG results. Starting from the Hamiltonian

$$H_{\text{nn}} = J_1 \sum_{\langle ij \rangle} \left( S_i^x S_j^x + S_i^y S_j^y + S_i^z S_j^z \right) , \tag{20}$$

we first perform a local rotation of the coordinate system in order to align the spins in the 120° state locally with the rotated $z'$-axis (spins are assumed to lie in the $x$-$z$-plane):

$$\begin{aligned} S_i^z &= \cos(\theta_i) \, S_i^{z'} - \sin(\theta_i) \, S_i^{x'} \\ S_i^x &= \sin(\theta_i) \, S_i^{z'} + \cos(\theta_i) \, S_i^{x'} . \end{aligned} \tag{21}$$

This yields

$$\begin{aligned} H_{\text{nn}} = J_1 \sum_{\langle ij \rangle} & \left[ S_i^{y'} S_j^{y'} + \cos(\theta_i - \theta_j) \left[ S_i^{z'} S_j^{z'} + S_i^{x'} S_j^{x'} \right] \right. \\ & \left. + \sin(\theta_i - \theta_j) \left[ S_i^{z'} S_j^{x'} - S_i^{x'} S_j^{z'} \right] \right] \end{aligned} \tag{22}$$

Applying the Holstein-Primakoff transformation $S_i^{z'} = S - a_i^\dagger a_i$ and $S_i^+ = \sqrt{2S - a_i^\dagger a_i} \, a_i$, we obtain in leading order

$$H_{\text{nn}} = H_0 + H_2 \tag{23}$$

with

$$H_0 = -\frac{3}{2} J_1 S^2 N \tag{24}$$

and the quadratic term

$$\begin{aligned} H_2 = J_1 S \sum_\mathbf{r} \sum_\delta & \left[ -\frac{3}{4} \left[ a_\mathbf{r} a_{\mathbf{r}+\delta} + a_{\mathbf{r}+\delta}^\dagger a_\mathbf{r}^\dagger \right] \right. \\ & + \frac{1}{4} \left[ a_{\mathbf{r}+\delta}^\dagger a_\mathbf{r} + a_\mathbf{r}^\dagger a_{\mathbf{r}+\delta} \right] + \frac{1}{2} \left[ a_\mathbf{r}^\dagger a_\mathbf{r} + a_{\mathbf{r}+\delta}^\dagger a_{\mathbf{r}+\delta} \right] \right] , \end{aligned} \tag{25}$$

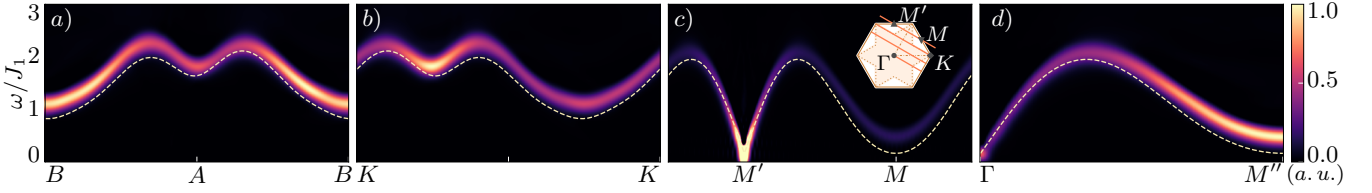

FIG. 6. Dynamical spin structure factor $S^{+-}(\mathbf{k}, \omega)$ in the stripe-ordered phase at $\frac{J_2}{J_1} = 0.55$. The simulations have been done on a cylinder of size $6 \times 126$ with bond dimensions $\chi = 800$. The dashed line denotes the dispersion from linear spin-wave theory (cf. Sec. **Linear Spin-Wave Theory in the Stripe-Ordered Phase** in this Supplemental Material).

where $\boldsymbol{\delta} = \mathbf{a}_1, \mathbf{a}_2, \mathbf{a}_3$ sums over half of the neighboring sites. The lattice translation vectors are $\mathbf{a}_1 = (1, 0)^T$, $\mathbf{a}_2 = \left(\frac{1}{2}, \frac{\sqrt{3}}{2}\right)^T$ and $\mathbf{a}_3 = \mathbf{a}_1 - \mathbf{a}_2$. Moreover, we used that for nearest-neighbor pairs, $\cos(\theta_i - \theta_j) = -\frac{1}{2}$ holds. Fourier transforming this expression results in

$$H_2 = \sum_{\mathbf{k}} \left[ A_{\mathbf{k}} a_{\mathbf{k}}^\dagger a_{\mathbf{k}} - \frac{1}{2} B_{\mathbf{k}} \left[ a_{\mathbf{k}}^\dagger a_{-\mathbf{k}}^\dagger + a_{-\mathbf{k}} a_{\mathbf{k}} \right] \right] \quad (26)$$

with $A_{\mathbf{k}} = 3 J_1 S \left( 1 + \frac{1}{2} \gamma_{\mathbf{k}} \right)$, $B_{\mathbf{k}} = \frac{9}{2} J_1 S \gamma_{\mathbf{k}}$ and

$$\gamma_{\mathbf{k}} = \frac{1}{6} \sum_{\boldsymbol{\delta}} \left( e^{i\mathbf{k}\cdot\boldsymbol{\delta}} + e^{-i\mathbf{k}\cdot\boldsymbol{\delta}} \right) \quad (27)$$

$$= \frac{1}{3} \left[ \cos(k_x) + 2 \cos\left(\frac{k_x}{2}\right) \cos\left(k_y \frac{\sqrt{3}}{2}\right) \right]. \quad (28)$$

For $\mathbf{k} \neq 0, \mathbf{K}, \mathbf{K}'$, this Hamiltonian can be diagonalized via a Bogoliubov transformation $a_{\mathbf{k}} = u_{\mathbf{k}} b_{\mathbf{k}} + v_{\mathbf{k}} b_{-\mathbf{k}}^\dagger$ with the coefficients being given as

$$u_{\mathbf{k}} = \left( \frac{A_{\mathbf{k}} + \varepsilon_{\mathbf{k}}}{2\varepsilon_{\mathbf{k}}} \right)^{\frac{1}{2}} \quad (29)$$

$$v_{\mathbf{k}} = \text{sgn}(B_{\mathbf{k}}) \left( \frac{A_{\mathbf{k}} - \varepsilon_{\mathbf{k}}}{2\varepsilon_{\mathbf{k}}} \right)^{\frac{1}{2}}. \quad (30)$$

and

$$\varepsilon_{\mathbf{k}} = \sqrt{A_{\mathbf{k}}^2 - B_{\mathbf{k}}^2} = 3 J_1 S \sqrt{(1 - \gamma_{\mathbf{k}})(1 + 2\gamma_{\mathbf{k}})}. \quad (31)$$

This leads to the expression

$$H_2 = \sum_{\mathbf{k}} \left[ \varepsilon_{\mathbf{k}} \left( b_{\mathbf{k}}^\dagger b_{\mathbf{k}} + \frac{1}{2} \right) - \frac{A_{\mathbf{k}}}{2} \right] \quad (32)$$

(note that $\varepsilon_{\mathbf{k}}$ vanishes for $\mathbf{k} = 0, \mathbf{K}, \mathbf{K}'$). We finally end up with the result [20]

$$H_{nn} = -\frac{3}{2} J_1 N S(S + 1) + \sum_{\mathbf{k}} \varepsilon_{\mathbf{k}} \left( b_{\mathbf{k}}^\dagger b_{\mathbf{k}} + \frac{1}{2} \right), \quad (33)$$

where we used the identity $\sum_{\mathbf{k}} \left( 1 + \frac{\gamma_{\mathbf{k}}}{2} \right) = L_y \cdot L_x \equiv N$.

*Static Structure Factor.*—We need to calculate the correlations between different sites $\mathbf{r}$ and $\mathbf{r}'$ on the lattice to compute the static structure factor $\chi(\mathbf{q})$ in (2) of the main text. For the classical 120° Néel state we find

$$\langle \mathbf{S_r} \cdot \mathbf{S_{r'}} \rangle_{\text{class}} = \begin{cases} \frac{3}{4} & \text{if } \mathbf{r} = \mathbf{r}' \\ \frac{1}{4} & \text{if } \mathbf{r}, \mathbf{r}' \text{ on the same sublattice} \\ -\frac{1}{8} & \text{if } \mathbf{r}, \mathbf{r}' \text{ on different sublattices} \end{cases}$$

We can treat quantum fluctuations within linear spin-wave theory. The spin components on site $i$ are given in terms of the bosonic operators as

$$S_i^{z'} = S - a_i^\dagger a_i \quad (34)$$

$$S_i^{x'} = \sqrt{\frac{S}{2}} \left( a_i + a_i^\dagger \right) \quad (35)$$

$$S_i^{y'} = \frac{1}{i} \sqrt{\frac{S}{2}} \left( a_i - a_i^\dagger \right). \quad (36)$$

We find the following correlations

$$\langle S_{\mathbf{r}}^y \cdot S_{\mathbf{r}'}^y \rangle = \frac{S}{2N} \sum_{\mathbf{k}}' \sqrt{\frac{A_{\mathbf{k}} - B_{\mathbf{k}}}{A_{\mathbf{k}} + B_{\mathbf{k}}}} e^{i\mathbf{k}\cdot(\mathbf{r}-\mathbf{r}')} \quad (37)$$

$$\langle S_{\mathbf{r}}^x \cdot S_{\mathbf{r}'}^x \rangle = \frac{S}{2N} \cos(\theta_{\mathbf{r}-\mathbf{r}'}) \sum_{\mathbf{k}}' \sqrt{\frac{A_{\mathbf{k}} + B_{\mathbf{k}}}{A_{\mathbf{k}} - B_{\mathbf{k}}}} e^{i\mathbf{k}\cdot(\mathbf{r}-\mathbf{r}')} \quad (38)$$

$$\langle S_{\mathbf{r}}^z \cdot S_{\mathbf{r}'}^z \rangle = \left\{ S^2 - 2 \frac{S}{N} \sum_{\mathbf{k}}' v_{\mathbf{k}}^2 + \frac{1}{N^2} \left( \sum_{\mathbf{k}}' v_{\mathbf{k}}^2 \right)^2 \right.$$
$$+ \frac{1}{N^2} \left( \sum_{\mathbf{k}}' u_{\mathbf{k}} v_{\mathbf{k}} e^{i\mathbf{k}\cdot(\mathbf{r}-\mathbf{r}')} \right)^2$$
$$\left. + \frac{1}{N^2} \sum_{\mathbf{k}}' \sum_{\mathbf{k}'}' v_{\mathbf{k}}^2 u_{\mathbf{k}'}^2 e^{i(\mathbf{k}-\mathbf{k}')\cdot(\mathbf{r}-\mathbf{r}')} \right\} \cos(\theta_{\mathbf{r}-\mathbf{r}'}) \quad (39)$$

with the angular contribution being given as $\cos(\theta_{\mathbf{r}-\mathbf{r}'}) = \cos\left( \frac{2\pi}{3} (n_1 - n_1' - (n_2 - n_2')) \right)$ for $\mathbf{r}^{(\prime)} = n_1^{(\prime)} \mathbf{a}_1 + n_2^{(\prime)} \mathbf{a}_2$. The primed sums denote momentum sums over all momenta in the Brillouin zone excluding the gapless Goldstone modes at $0$, $K$ and $K'$. The momenta can be written as $\mathbf{k} = \frac{m_1}{L_x} \mathbf{b}_1 + \frac{m_2}{L} \mathbf{b}_2$ ($m_i \in \mathbb{Z}$). Note that the corners of the Brillouin zone $K$ and $K'$ are included if and only if $L_x$ and $L_y$ are multiples of 3.

**Linear Spin-Wave Theory in the Stripe-Ordered Phase**

We can apply a similar formalism as in the previous paragraph to obtain the magnon dispersion resulting from lowest-order spin-wave theory in the stripe-ordered phase at $\frac{J_2}{J_1} = 0.55$ (cf. Fig. 6). The standard approach for an antiferromagnetic ground state introduces two different kinds of bosonic operators on the sublattices of the system with opposite spin orientation [21, 22]. We want to apply a different approach in analogy to the one presented above by locally rotating the reference frame. We consider the full Hamiltonian with nearest- and next-nearest-neighbor interaction:

$$H = J_1 \sum_{\langle ij \rangle} \left( S_i^x S_j^x + S_i^y S_j^y + S_i^z S_j^z \right)$$
$$+ J_2 \sum_{\langle\langle ij \rangle\rangle} \left( S_i^x S_j^x + S_i^y S_j^y + S_i^z S_j^z \right) \quad (40)$$

We rotate the local frame according to equation (21) by an angle of $\pi$. The orientation of the spins in the ground state hereby alternates when going into the direction of $\mathbf{a}_2$. This assumption can be verified from the numerical ground state found by DMRG at the coupling $\frac{J_2}{J_1} = 0.55$.

The nearest-neighbor pairs are given for each site as before by $\boldsymbol{\delta}_{nn}^i = \mathbf{a}_i$, $i = 1, 2, 3$. For the next-nearest-neighbor pairs we choose $\boldsymbol{\delta}_{nnn}^i = \mathbf{l}_i$ ($i = 1, 2, 3$) with

$$\mathbf{l}_1 = 2\mathbf{a}_2 - \mathbf{a}_1$$
$$\mathbf{l}_2 = \mathbf{a}_1 + \mathbf{a}_2$$
$$\mathbf{l}_3 = 2\mathbf{a}_1 - \mathbf{a}_2 \,. \quad (41)$$

Defining $\Delta\theta(\mathbf{d}) = \theta_{\mathbf{r}} - \theta_{\mathbf{r+d}}$ we obtain

$$\Delta\theta\left(\mathbf{a}_1\right) = \Delta\theta\left(\mathbf{l}_1\right) = 0 \quad (42)$$
$$\Delta\theta\left(\mathbf{a}_{2/3}\right) = \Delta\theta\left(\mathbf{l}_{2/3}\right) = \pm\pi \,. \quad (43)$$

resulting in $\cos(\Delta\theta) \in \{1, -1\}$ for the different cases.

After applying the standard Holstein-Primakoff transformation as introduced in the previous paragraph, the zeroth order Hamiltonian yields

$$H_0 = -S^2 \left( J_1 + J_2 \right) N \,. \quad (44)$$

The second order contribution reads

$$H_2 = J_1 S \sum_{\langle ij \rangle} \Bigg\{ -\cos\left(\theta_i - \theta_j\right) \left( a_i^\dagger a_i + a_j^\dagger a_j \right)$$
$$+ \left( a_i a_j + a_j^\dagger a_i^\dagger \right) \frac{\cos\left(\theta_i - \theta_j\right) - 1}{2}$$
$$+ \left( a_i^\dagger a_j + a_j^\dagger a_i \right) \cdot \frac{\cos\left(\theta_i - \theta_j\right) + 1}{2} \Bigg\}$$
$$+ J_2 S \sum_{\langle\langle ij \rangle\rangle} \ldots \quad (45)$$

The terms for the next-nearest-neighbor contribution have the same form as the nearest-neighbor ones.

A transformation to Fourier space results in a Hamiltonian with a similar structure as in equation (26):

$$H_2 = \sum_{\mathbf{k}} \left[ \tilde{A}_{\mathbf{k}} a_{\mathbf{k}}^\dagger a_{\mathbf{k}} - \frac{1}{2} \left( \tilde{B}_{\mathbf{k}}^* a_{\mathbf{k}}^\dagger a_{-\mathbf{k}}^\dagger + \tilde{B}_{\mathbf{k}} \, a_{-\mathbf{k}} a_{\mathbf{k}} \right) \right] \quad (46)$$

with

$$\tilde{A}_{\mathbf{k}} = 2J_1 S \left[ 1 + \cos\left(\mathbf{k} \cdot \mathbf{a}_1\right) \right]$$
$$+ 2J_2 S \left[ 1 + \cos\left(\mathbf{k} \cdot \mathbf{l}_1\right) \right] \quad (47)$$

and

$$\tilde{B}_{\mathbf{k}} = 2J_1 S \left[ e^{i\mathbf{k}\cdot\mathbf{a}_2} + e^{i\mathbf{k}\cdot\mathbf{a}_3} \right]$$
$$+ 2J_2 S \left[ e^{i\mathbf{k}\cdot\mathbf{l}_2} + e^{i\mathbf{k}\cdot\mathbf{l}_3} \right] \,. \quad (48)$$

One can diagonalize this Hamiltonian again via a canonical Bogoliubov transformation $a_{\mathbf{k}} = u_{\mathbf{k}} b_{\mathbf{k}} + v_{\mathbf{k}} b_{-\mathbf{k}}^\dagger$ with

$$u_{\mathbf{k}} = \left( \frac{\tilde{A}_{\mathbf{k}} + \varepsilon_{\mathbf{k}}}{2\varepsilon_{\mathbf{k}}} \right)^{\frac{1}{2}} \quad (49)$$

and

$$v_{\mathbf{k}} = \mathrm{sgn}\left( \mathrm{Re}\{\tilde{B}_{\mathbf{k}}\} \right) \left( \frac{\tilde{A}_{\mathbf{k}} - \varepsilon_{\mathbf{k}}}{2\varepsilon_{\mathbf{k}}} \right)^{\frac{1}{2}} , \quad (50)$$

yielding

$$H_2 = \sum_{\mathbf{k}} \left[ \varepsilon_{\mathbf{k}} \left( b_{\mathbf{k}}^\dagger b_{\mathbf{k}} + \frac{1}{2} \right) - \frac{\tilde{A}_{\mathbf{k}}}{2} \right] \,. \quad (51)$$

The energy dispersion is given as

$$\varepsilon_{\mathbf{k}} = \sqrt{\tilde{A}_{\mathbf{k}}^2 - \left( \mathrm{Re}\{\tilde{B}_{\mathbf{k}}\} \right)^2} \,. \quad (52)$$

**F: Mean-Field Theory for the Dirac Spin Liquid**

Besides spin-wave theory, we explore parton mean-field theory [23–25] in the QSL phase to compare to our numerics. Starting from the Hamiltonian in equation (1) in the main text, one can substitute the spin operators by fermionic spinon operators $c_{i\alpha}$, $c_{i\alpha}^\dagger$ (the Greek letters run over the spin labels $\uparrow, \downarrow$) using the common parton construction

$$\mathbf{S}_i = \frac{1}{2} c_{i\alpha}^\dagger \boldsymbol{\sigma}_{\alpha\beta} c_{i\beta} \,, \quad (53)$$

where $\boldsymbol{\sigma} = (\sigma_x, \sigma_y, \sigma_z)$ contains the Pauli matrices. In order to describe the field theory for a Dirac spin liquid,

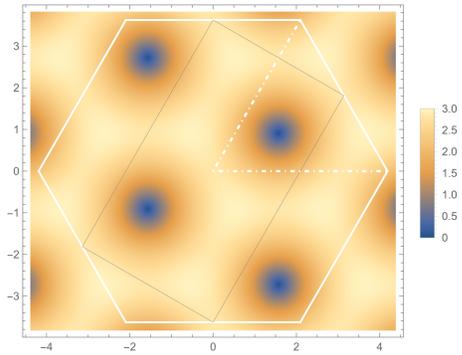

FIG. 7. Dispersion $\varepsilon_{\mathbf{k}}$ of the upper spinon band at half-filling (equation (61)) for the mean-field Dirac spin liquid. The white solid lines mark the Brillouin zone of the lattice and the thin gray line the reduced first Brillouin zone of the spinons.

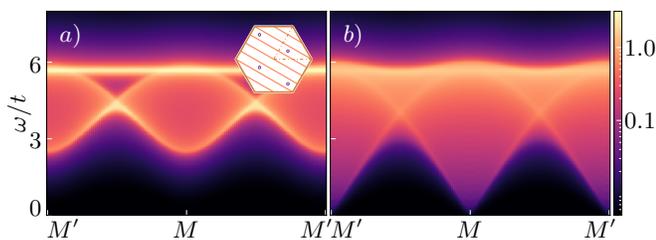

FIG. 8. Dynamical spin structure factor obtained from the mean-field parton approach for a Dirac spin liquid on a cylinder with $a)$ $L_y = 6$ and $b)$ $L_y = 100$ ($L_x = 142$ each). The inset in $a)$ shows the accessible momenta on a $L_y = 6$ cylinder as orange cuts (cf. Fig. 10). The gapless Dirac points in the spinon dispersion—denoted by violet dots—are not hit, which results in a gapped dispersion.

we keep only the spin-preserving hopping terms in the mean-field expansion

$$H_{\mathrm{MF}} = H_\uparrow + H_\downarrow \qquad (54)$$

$$H_\alpha = -\sum_{\langle i,j \rangle} \left( t_{ij} c_{i\alpha}^\dagger c_{j\alpha} + \mathrm{h.c.} \right) \qquad (55)$$

and choose the phase of the hopping constant $t_{ij} = t\, e^{\mathrm{i}\varphi_{ij}}$ such that the triangles of the lattice have alternatingly flux $\pi$ and 0. This flux pattern requires a two-site unit cell. The Bravais vectors are then $2\mathbf{a}_1$ and $\mathbf{a}_2$, whereas the reciprocal lattice vectors become $\frac{\mathbf{b}_1}{2}$ and $\mathbf{b}_2$. Defining the Fourier transformation accordingly

$$c_{(n_1, n_2, \delta)\alpha} = \frac{1}{\sqrt{L_y \frac{L_x}{2}}} \sum_{\mathbf{k}} e^{\mathrm{i}\mathbf{k}\cdot\mathbf{r}_{n_1, n_2, \delta}} c_{\mathbf{k}\delta\alpha}, \qquad (56)$$

where $n_1$, $n_2$ label the unit cell and $\delta = 0, 1$ the site within the unit cell ($i \equiv (n_1, n_2, \delta)$, i.e., $\mathbf{r} = \mathbf{r}_i = (2\,n_1 + \delta)\,\mathbf{a}_1 + n_2\,\mathbf{a}_2$), we can bring the Hamiltonian into the form

$$H_\alpha = -t \sum_{\mathbf{k}} \left( c_{\mathbf{k}0\alpha}^\dagger c_{\mathbf{k}1\alpha}^\dagger \right) \mathcal{H} \begin{pmatrix} c_{\mathbf{k}0\alpha} \\ c_{\mathbf{k}1\alpha} \end{pmatrix} \qquad (57)$$

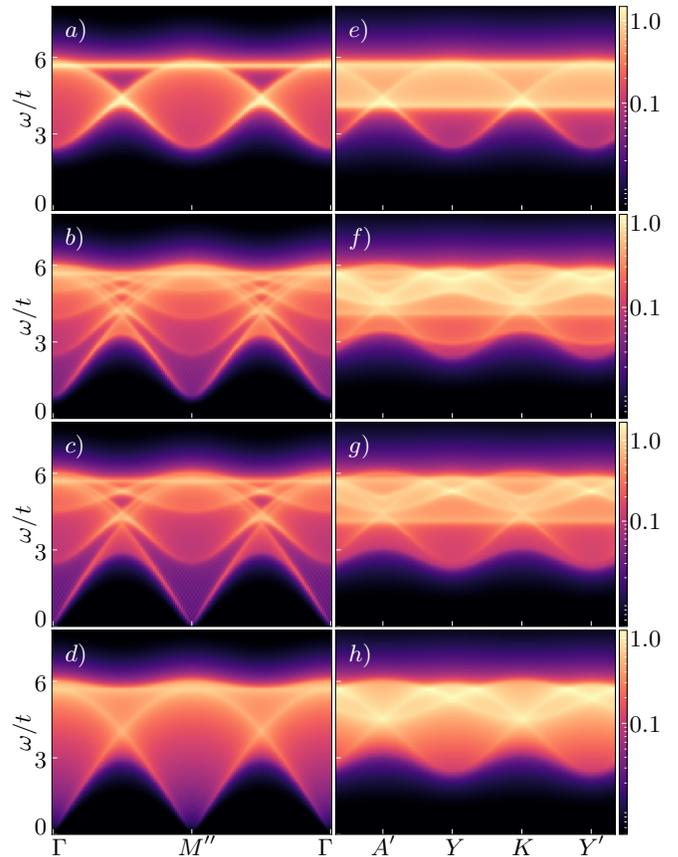

FIG. 9. Dynamical spin structure factor from the mean-field spinon theory on different cylinder circumferences: $a)$, $e)$ $L_y = 6$, $b)$, $f)$ $L_y = 9$, $c)$, $g)$ $L_y = 12$, $d)$ $L_y = 100$ and $h)$ $L_y = 96$. We apply a logarithmic color scale. See Fig. 10 for the labels of the momentum points. All calculations have been done for $L_x = 142$.

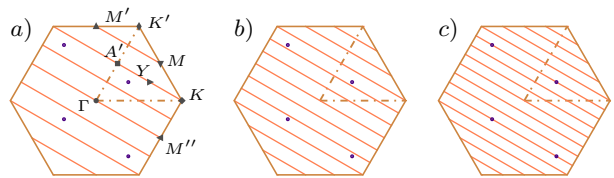

FIG. 10. Accessible momenta on cylinders with circumferences $a)$ $L_y = 6$, $b)$ $L_y = 9$ and $c)$ $L_y = 12$ for $L_x \to \infty$. The violet dots denote the gapless Dirac points $\pm\mathbf{Q}$ of the spinon theory and equivalent momentum points. Note that the first Brillouin zone for the spinons has only half the size of the one for the spin model.

with

$$\mathcal{H} = \begin{pmatrix} 2\cos(\mathbf{k} \cdot \mathbf{a}_2) & \kappa(\mathbf{k}) \\ \kappa^*(\mathbf{k}) & -2\cos(\mathbf{k} \cdot \mathbf{a}_2) \end{pmatrix} \qquad (58)$$

and

$$\kappa_{\mathbf{k}} = 2\cos(\mathbf{k} \cdot \mathbf{a}_1) + 2\mathrm{i}\sin(\mathbf{k} \cdot (\mathbf{a}_1 - \mathbf{a}_2)) . \qquad (59)$$

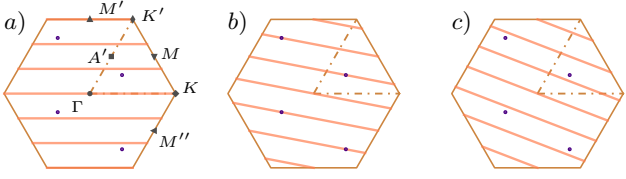

FIG. 11. Accessible momenta on cylinders with $L = 6$ for different geometries $a)$ YC6 − 3, $b)$ YC6 − 2 and $c)$ YC6 − 1 for $L_x \to \infty$. The violet dots denote the gapless Dirac points $\pm \mathbf{Q}$ of the spinon theory and equivalent momentum points. The geometry YC$L - n$ is defined in such a way that the periodicity along the circumference of the cylinder is closed by the translation $L \cdot \mathbf{a}_2 - n \cdot \mathbf{a}_1$ (cf. [4]).

The resulting energy is

$$\varepsilon_{\mathbf{k}} = \sqrt{4 \cos^2 (\mathbf{k} \cdot \mathbf{a}_2) + |\kappa_{\mathbf{k}}|^2} \tag{60}$$

$$= \sqrt{2} \sqrt{3 + \cos(2k_x) - 2 \sin(k_x) \sin(\sqrt{3}k_y)} \,. \tag{61}$$

The diagonal Hamiltonian reads

$$H_\alpha = t \sum_{\mathbf{k}} \varepsilon_{\mathbf{k}} \left( \gamma_{\mathbf{k}1\alpha}^\dagger \gamma_{\mathbf{k}1\alpha} - \gamma_{\mathbf{k}0\alpha}^\dagger \gamma_{\mathbf{k}0\alpha} \right) \tag{62}$$

with the transformed operators given as

$$\gamma_{\mathbf{k}0\alpha} = \frac{\kappa^*}{\sqrt{2\varepsilon}\sqrt{\varepsilon - a}} c_{\mathbf{k}0\alpha} + \sqrt{\frac{\varepsilon - a}{2\varepsilon}} c_{\mathbf{k}1\alpha} \tag{63}$$

$$\gamma_{\mathbf{k}1\alpha} = \frac{\kappa^*}{\sqrt{2\varepsilon}\sqrt{\varepsilon + a}} c_{\mathbf{k}0\alpha} - \sqrt{\frac{\varepsilon + a}{2\varepsilon}} c_{\mathbf{k}1\alpha} \,. \tag{64}$$

Note that we suppressed the momentum label $\mathbf{k}$ in $\kappa_{\mathbf{k}}$, $\varepsilon_{\mathbf{k}}$ and $a_{\mathbf{k}} = 2 \cos(\mathbf{k} \cdot \mathbf{a}_2)$. At half-filling, the lower band is occupied and Dirac cones occur at $\pm \mathbf{Q} = \pm \left( \frac{\pi}{2}, \frac{\pi}{2\sqrt{3}} \right)$. For the static spatial correlations, we find for all spin components $\beta = x, y, z$:

$$\langle S_{\mathbf{r}}^\beta S_{\mathbf{r}'}^\beta \rangle = \frac{2}{(L_x L_y)^2} \sum_{\mathbf{k},\mathbf{q}} e^{-i(\mathbf{k}-\mathbf{q}) \cdot (\mathbf{r} - \mathbf{r}')} g_{\delta\delta'}(\mathbf{k}, \mathbf{q}). \tag{65}$$

For the sake of notational convenience, we introduced here the function

$$g_{\delta\delta'}(\mathbf{k}, \mathbf{q}) := \begin{cases} \frac{|\kappa|^2}{2\varepsilon(\varepsilon - a)}\Big|_{\mathbf{k}} \cdot \frac{|\kappa|^2}{2\varepsilon(\varepsilon + a)}\Big|_{\mathbf{q}} & \text{if } \delta = \delta' = 0 \\ \frac{\varepsilon - a}{2\varepsilon}\Big|_{\mathbf{k}} \cdot \frac{\varepsilon + a}{2\varepsilon}\Big|_{\mathbf{q}} & \text{if } \delta = \delta' = 1 \\ -\frac{\kappa^*}{2\varepsilon}\Big|_{\mathbf{k}} \cdot \frac{\kappa}{2\varepsilon}\Big|_{\mathbf{q}} & \text{if } \delta = 0, \delta' = 1 \\ -\frac{\kappa}{2\varepsilon}\Big|_{\mathbf{k}} \cdot \frac{\kappa^*}{2\varepsilon}\Big|_{\mathbf{q}} & \text{if } \delta = 1, \delta' = 0 \end{cases} \tag{66}$$

Hence, the complete static correlation function at equal time is

$$\langle \mathbf{S}_{\mathbf{r}} \cdot \mathbf{S}_{\mathbf{r}'} \rangle = 3 \cdot \langle S_{\mathbf{r}}^x S_{\mathbf{r}'}^x \rangle \,. \tag{67}$$

The static spin structure factor follows straightforwardly.

*Dynamic Spin Structure Factor.*—We can rewrite the momentum-dependent ladder operator

$$S_{\mathbf{k}}^- = \frac{1}{\sqrt{N}} \sum_{\mathbf{r}} e^{i\mathbf{k}\cdot\mathbf{r}} S_{\mathbf{r}}^- \tag{68}$$

in terms of the fermionic spinon operators:

$$S_{\mathbf{k}}^- = \sum_\delta \sum_{\mathbf{q}} c_{\mathbf{q}+\mathbf{k}\delta\downarrow}^\dagger c_{\mathbf{q}\delta\uparrow} \,. \tag{69}$$

The dynamical spin structure factor is defined as

$$S(\mathbf{k}, \omega) = \frac{1}{N} \int \mathrm{d}t\, e^{i\omega t} \sum_{\mathbf{r},\mathbf{r}'} e^{-i\mathbf{k}\cdot(\mathbf{r}-\mathbf{r}')} \langle S_{\mathbf{r}}^+(t) S_{\mathbf{r}'}^-(0) \rangle \tag{70}$$

Using the trivial time dependence of the operators $\gamma_{\mathbf{k}\delta\alpha}$, we obtain for the dynamical structure factor

$$S(\mathbf{k}, \omega) = 2\pi \sum_{\mathbf{q}} \delta \left( \omega - \varepsilon_{\mathbf{q}} - \varepsilon_{\mathbf{q}+\mathbf{k}} \right) \left| \langle \Phi_{\mathbf{k}}^{\mathbf{q}} | S_{\mathbf{k}}^- | \Psi_0 \rangle \right|^2 \,, \tag{71}$$

where the overlap with the excited state $|\Phi_{\mathbf{k}}^{\mathbf{q}}\rangle$ with quantum number $\mathbf{q}$ is given as

$$\left| \langle \Phi_{\mathbf{k}}^{\mathbf{q}} | S_{\mathbf{k}}^- | \Psi_0 \rangle \right|^2 = \frac{|\kappa_{\mathbf{q}}|^2 |\kappa_{\mathbf{q}+\mathbf{k}}|^2}{4\varepsilon_{\mathbf{q}}\varepsilon_{\mathbf{q}+\mathbf{k}}} \cdot \frac{1}{\varepsilon_{\mathbf{q}} - a_{\mathbf{q}}} \cdot \frac{1}{\varepsilon_{\mathbf{q}+\mathbf{k}} + a_{\mathbf{q}+\mathbf{k}}}$$
$$+ \frac{\varepsilon_{\mathbf{q}} - a_{\mathbf{q}}}{2\varepsilon_{\mathbf{q}}} \cdot \frac{\varepsilon_{\mathbf{q}+\mathbf{k}} + a_{\mathbf{q}+\mathbf{k}}}{2\varepsilon_{\mathbf{q}+\mathbf{k}}} \tag{72}$$

Here, $|\Psi_0\rangle$ denotes the spinon ground state at half-filling. For the purpose of plotting, we model the delta-distribution in Figs. 8 and 9 as a Lorentzian

$$\delta(\omega - \varepsilon) \to \frac{\eta/\pi}{(\omega - \varepsilon)^2 + \eta^2} \,, \tag{73}$$

where we chose a broadening of $\eta = 0.1\, t$ (cf., e.g., [26]).

The resulting dynamic spin structure factor for different cylinder circumferences $L_y$ obtained from spinon mean-field theory are shown in Fig. 9. Note that the constraint $\sum_\alpha c_{i\alpha}^\dagger c_{i\alpha} = 1$ is only satisfied on average in the mean-field treatment. (We do not apply here a Gutzwiller projector for the spectral functions shown in Figs. 8 and 9). We observe that the minima at the $M$ points depend strongly on the choice of $L_y$ for small cylinder circumferences. While for $L_y = 12$ it is gapless, there is a clear gap for $L_y = 6$. This is caused by the restrictions that the value of $L_y$ imposes on the accessible momenta. For $L_y = 6$, these do not include the gapless Dirac points at $\pm \mathbf{Q}$ (cf. Fig. 10). For $L_y = 9$, momenta closer to the Dirac points are available, whereas for $L_y = 12$, one cut hits directly the points $\pm \mathbf{Q}$ each. This is in line with the arguments given in [27]. Note that the minima in the dispersion at the corners of the Brillouin zone are not captured in the mean-field spinon theory. To resolve them in the variational Monte Carlo simulations, a Gutzwiller projection of the wave function is required [28].

## G: Different Cylinder Geometries

The accessible momenta in the Brillouin zone also depend on the geometry of the cylinder, by which term we mean how to close the periodic circumference of the cylinder. Following the notation of Hu et al. [4], we define the geometry YC$L - n$ by the periodic translation vector $L \cdot \mathbf{a}_2 - n \cdot \mathbf{a}_1$. The standard way to close the periodicity that we chose for the simulations presented in this work is YC$L - 0 \equiv$ YC$L$. The accessible parts of the Brillouin zone are shown in Fig. 10 for various $L$. Other possible geometries that come with a similar computational complexity for a given $L$ since the resulting matrix product operator representing the Hamiltonian has comparable virtual bond dimensions are shown in Fig. 11. The $J_1 - J_2$ Heisenberg model on the triangular lattice for $L = 6$ as applied for the simulations in this work has for instance a virtual bond dimension of $\chi_{\mathrm{MPO}} = 38$ if $J_2 \neq 0$. The advantage of the chosen geometry YC$L$, as apparent from Figs. 10 and 11, is that all high-symmetry momenta in the Brillouin zone such as the corners and the midpoints of the edges of the Brillouin zone are accessible.

---

* markus.drescher@tum.de



\end{document}